# Design of Self-Folding Origami Architectures for Biomedical Implants: A Finite Element Simulation Framework

Suhas Chandrashekar[1], Thomas Stieglitz[1], Prasanth Velvaluri*[1]

[1] Laboratory for Biomedical Microtechnology, Department of Microsystem Engineering (IMTEK), IMBIT//Neuro Probes, Brain Links-Brain Tools, University of Freiburg, Germany

*Corresponding Author

## Abstract:

Minimally invasive brain implants require compact deployment geometries and controlled shape transformation for delivery through confined vessels. Origami-inspired structures enable large shape changes from compact to tailored deployed configurations, but implementation at microscale dimensions remains challenging, as conventional designs rely on predefined mountain and valley folds that become increasingly difficult to fabricate and assemble as unit-cell size is reduced to meet clinical requirements. To address this, we propose a self-folding methodology based on thermally induced stresses in thin-film bimetallic composites. Pairing the shape memory alloy Nitinol (NiTi) with molybdenum or stainless steel (SS316L) exploits their differing thermal expansion coefficients to generate localized hinge bending, enabling deterministic, autonomous folding into prescribed mountain and valley configurations. Using finite element simulations, we systematically optimize hinge length, layer thickness, width, and shape, identifying a hinge geometry achieving a ~50% reduction in unit-cell dimensions relative to previously reported designs. Integration into a complete waterbomb origami unit cell confirms that the strip-level folding mechanism scales to produce coordinated, self-folding deformation across the tessellated structure. These results demonstrate the feasibility of stress-induced, self-folding thin-film origami architectures built from clinically established biomaterials, providing a scalable design framework for next-generation, active neurovascular implants with integrated sensing and monitoring functionality.

# 1. Introduction

Intracranial aneurysms [1] and ischemic stroke [2] are among the leading causes of neurological disability and mortality worldwide. Endovascular treatment strategies have become the preferred clinical approach for many neurovascular diseases due to their minimally invasive nature and reduced patient recovery times compared with open surgical procedures [3].

Stents are widely used in neurovascular interventions to restore or modify blood flow within diseased vessels. In aneurysm treatment, flow-diverting stents reduce the inflow of blood into the aneurysm sac, thereby promoting thrombosis and vessel remodeling [4]. In ischemic stroke treatment, stents can restore perfusion by reopening occluded vessels and maintaining vessel patency [5]. Over the past two decades, numerous stent architectures have been developed, including braided stents, laser-cut stents, and flow-diversion devices [6,7]. The majority of these devices are fabricated using shape memory alloys (SMAs) that allow self-expansion of the stent due to superelasticity [8].

Despite their clinical success, current stent technologies largely rely on passive mechanical functionality. Their geometric architectures are primarily optimized for deployment, radial support, and flow modification, while providing limited opportunities for the integration of additional sensing, actuation, or communication functionalities [9]. As a result, there remains a growing interest in the development of next-generation smart stents capable of providing treatment feedback, monitoring local physiological conditions, or actively adapting to their surrounding environment [10].

Origami-inspired engineering has emerged as a promising design paradigm for deployable biomedical devices due to its ability to transform planar structures into complex three-dimensional geometries [11]. Origami-based concepts have been investigated for applications including surgical tools, drug-delivery systems, tissue scaffolds, micro actuators[12], and deployable implants [13]. In particular, origami-inspired stents offer several potential advantages over conventional designs, including compact packaging, programmable shape transition after deployment, and the ability to realize patient-specific geometries [14,15].

The realization of origami-inspired implants at microscale dimensions has been facilitated by advances in microsystem technologies and thin-film fabrication techniques [15,16]. Among the available smart materials, binary NiTi based shape memory alloys (SMA) are particularly attractive because they can be biocompatible, allow high design freedom, and generate large recoverable strains with thin-film processing approaches [17,18].

Recent proof-of-concept studies have demonstrated the feasibility of origami-inspired SMA stents fabricated using microsystem technologies [16]. These structures can be folded into compact configurations for minimally invasive delivery while maintaining designated low-stress regions that may serve as integration platforms for future sensing or electronic components. This can ease the flexibility requirements for the electronics in the smart stents, which is one of the key challenges. However, the relatively large unit-cell dimensions investigated thus far limit their applicability for neurovascular implantation. For practical catheter-based deployment in cerebral vessels, a substantial reduction in unit-cell dimensions is required.

While microsystem fabrication technologies readily enable geometric miniaturization, achieving the prescribed mountain and valley crease patterns becomes increasingly challenging as structural dimensions decrease. Conventional manual folding approaches become impractical at these scales, necessitating the development of integrated self-folding mechanisms capable of generating deterministic folding directions[19]. The ability to reliably initiate the desired crease orientation therefore represents a key challenge in the realization of miniaturized origami implants. Certain polymer-based origami structures have been demonstrated that trigger the self-folding based on changes in pH [15], temperature [20] or light [21]. However, the deployment mechanism and radial forces required for vascular stents limit the clinical use of such polymer-based devices.

To address this challenge, this work investigates the integration of thin-film metallic SMA-based bi-metal actuators along the origami crease lines (Figure 1). By exploiting thermally induced mismatch stresses within the actuator stack, localized bending moments are generated that initiate folding in a predefined direction [22]. The resulting deformation provides the initial mountain (SMA/Mo) or valley crease (SMA/SS316L) required for subsequent origami folding, both the Mo [23] and SS316L [24] are biocompatible. The effectiveness of this approach depends strongly on the geometric design of both the origami unit cell and the integrated actuator architecture.

In this paper, we systematically investigate the design space of miniaturized origami unit cells with dimensions reduced by approximately 50 % compared with previously reported experimental designs [16]. Using coupled Multiphysics simulations in COMSOL, the influence of key geometric parameters on self-folding performance is evaluated. The study identifies design guidelines for achieving reliable crease formation through SMA-driven bimetallic actuation and provides insight into the scaling limits of self-folding origami structures for future neurovascular implant applications.

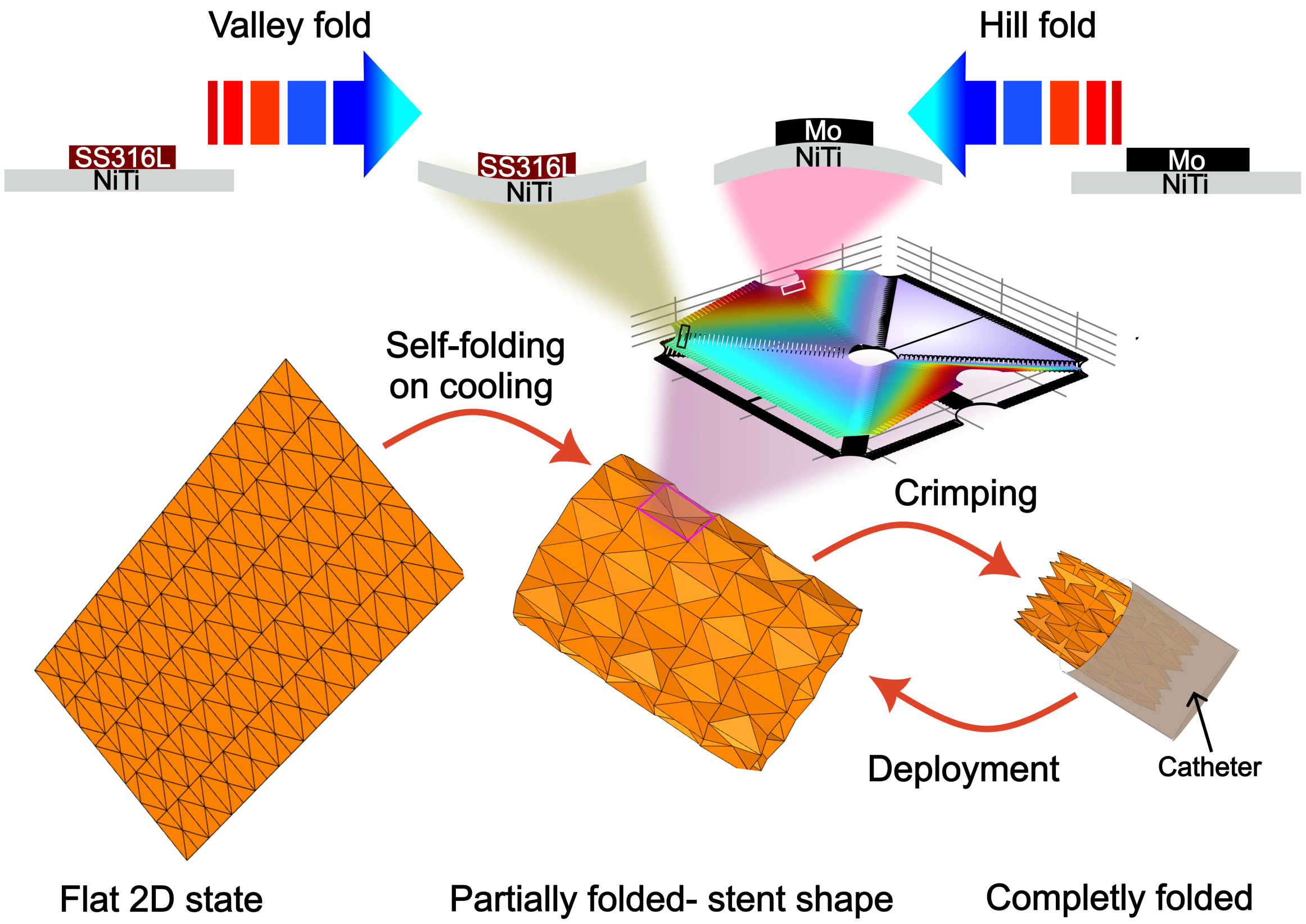

*Figure 1 Illustrated here is the concept of self-folding origami driven by shape memory alloy (SMA) actuators. Bilayer SMA/SS316L and SMA/Mo hinges, patterned along the valley and mountain creases respectively, undergo differential thermal contraction on cooling, driving localized bending at each hinge and progressively folding the initially flat crease pattern into the final stent geometry. The folded stent can subsequently be crimped using a conventional crimping tool to fit within a delivery catheter; upon deployment, the superelasticity of the SMA drives shape recovery to the deployed configuration. The waterbomb crease pattern was generated using Origami Simulator, an open-source origami-folding simulation tool [25] .*

# 2. Results and Discussion

## 2.1 Materials and geometry

The geometry of the actuator system is adapted from the origami based thin film shape memory alloy geometry structure [16], a so-called waterbomb origami pattern, where each unit cell is tessellated and connected to neighboring cells through shared hinges. Although in Figure 2a, only three such structures are shown, in reality such a device consists of a significantly large array of periodic repetition of the same unit cell.

To obtain a self-folding origami structure, the hinges of each unit have to deform to provide either a hill or a valley fold. A single unit cell (Figure 2b) consists of multiple hinges and each hinge has to either deform either to a hill (red) or a valley fold (blue). To enable self-folding the hinge was considered as a bimetallic actuator and each hinge consists of an array of strips.

All the strips work collectively and behave as an individual thermo-mechanical actuator during deformation. The collective interaction between the alternating mountain and valley folds throughout the array and the hinge results in overall folding motion of the unit cells. The repeating nature of the unit cell geometry results in scalable expansion of the structure while maintaining similar consistent local deformation characteristics.

The geometry of the unit cell is defined using several key design parameters like: strip length, layer thickness, strip width, overhang width and inter-strip width. These parameters were systematically optimized throughout the study to build a geometry capable of maximum controlled deformation.

**Strip Length ($L_{strip}$):** The strip length ($L_{strip}$) is defined as the longitudinal distance between the two ends of an individual actuator strip (Figure 2c). The strip length determines the length over which the thermo-mechanical actuation develops and is one of the principal geometric parameters.

**Layer Thickness ($T_{Mo/SS316L}$ and $T_{NiTi}$):** The strip is modelled as a bilayer structure consisting of NiTi bottom layer (grey) and top layer of metal either Molybdenum (Mo) or stainless steel (SS316L) based on the fold type (Figure 2d). $T_{Mo}$, $T_{SS316L}$ and $T_{NiTi}$ represent the thickness of molybdenum, stainless steel and Nitinol respectively. The mismatch in the thermal coefficient of expansion between these layers results in thermo-mechanical actuation.

**Strip width ($W_{strip}$):** The strip width ($W_{strip}$) is defined as the lateral distance of the actuator strip measured perpendicular to the $L_{strip}$ (Figure 2d). It represents the width of the functional metallic layer Mo or SS316L on top of NiTi and it defines the overall footprint of each actuator.

**Overhang Width ($W_{overhang}$):** The overhang width ($W_{overhang}$) is defined as the additional lateral extension of the bottom NiTi layer beyond the width of the top metal layer (Mo/SS316L) (Figure 2d). This overhang is essential in the fabrication aspect and also helps in reliable adhesion of the top layers to the NiTi layer. An overhang width of 1 µm was chosen and used for the current work, for details see supplementary 1.1 (Figure S1 and S2).

**Inter Strip Width ($W_{gap}$):** The inter strip width ($W_{gap}$) is defined as the spacing between adjacent strips (Figure 2d). This gap helps prevent localized stress concentration and also prevents the undesired lateral deflection and constrains the preferential longitudinal deflection. An interstrip width of 5 µm was chosen for this analysis, for details see supplementary 1.2 (Figure S3 and S4).

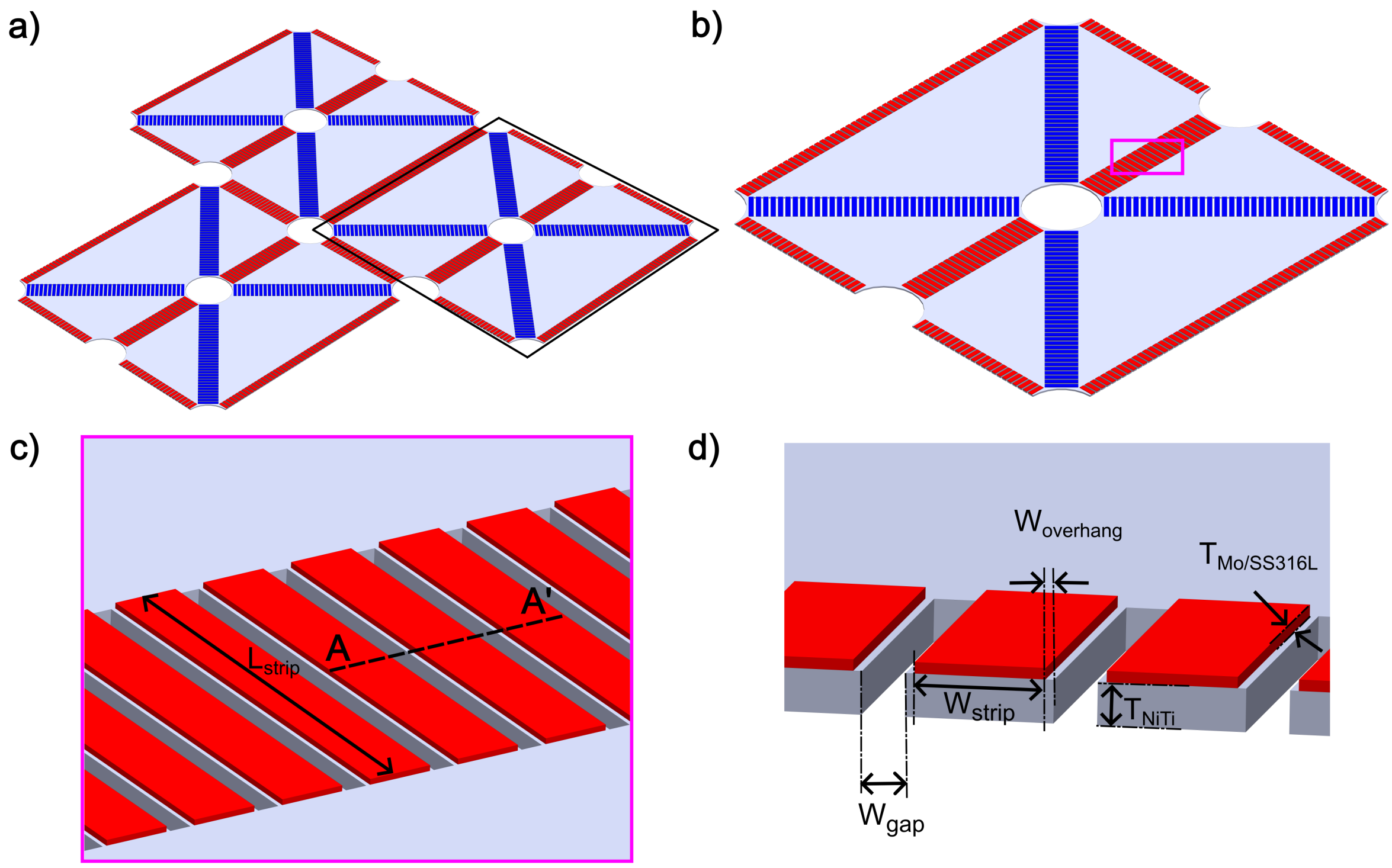

*Figure 2: Geometry of the waterbomb based origami self-folding actuator structure shown in 2D. (a) Representative waterbomb origami tessellated arrangement consisting of interconnected unit cells. (b) Single unit-cell geometry showing mountain (red) and valley (blue) fold actuator strips. (c) Represents the zoomed view of a hill hinge, definition of strip length ($L_{strip}$) of the actuator. (d) Represents in the cross-section AA' along the hill hinge, definition of geometric parameters strip width ($W_{strip}$), inter-strip gap ($W_{gap}$), overhang width ($W_{overhang}$), and layer thicknesses ($T_{Mo/SS316L}$, $T_{NiTi}$).*

## 2.2 Material considerations

For the present simulations, NiTi was approximated as a linear elastic material with austenitic phase material properties. This simplification was adopted because the primary objective of the study is to investigate the thermally induced mismatch stresses that generate the initial out-of-plane deformation required for self-folding. In the proposed NiTi–metal bilayer structures (NiTi/Mo and NiTi/SS316L), the dominant deformation mechanism originates from differential thermal contraction during cooling from the fabrication temperature (600 °C) to room temperature.

The NiTi composition considered in this work corresponds to a medical grade alloy (Ti-49 at. % Ni-51 at. %). For such compositions, the austenite finish temperature is typically below room temperature (~ 288 K), while the martensitic transformation temperature lies well below ambient conditions (~ 213 K) [26,27]. Consequently, throughout the temperature range considered in the simulations (873 K-293 K), the material remains in the austenitic phase and does not undergo a thermally induced phase transformation. Under these conditions, the deformation behavior is governed primarily by the mismatch in the coefficients of thermal expansion between the constituent layers rather than by transformation-induced strains. Therefore, a linear elastic approximation provides a reasonable representation of the physical mechanism responsible for crease initiation and self-folding.

It should be noted that the objective of the simulations is not to capture the full constitutive response of NiTi, but rather to evaluate the influence of geometric design parameters on the magnitude and direction of thermally induced bending. The use of a linear material model significantly reduces computational complexity while preserving the dominant physics associated with thermal mismatch actuation.

Although the shape memory and superelastic characteristics of NiTi were not explicitly considered in the present simulations, these properties remain essential for the functionality of the final device. The thermally induced mismatch stresses investigated here provide the initial self-folding mechanism required to establish the prescribed mountain and valley crease patterns. Following deployment, however, the superelastic behavior of NiTi is expected to facilitate mechanical unfolding and recovery of the origami structure without permanent deformation [16]. Consequently, the current study focuses on the geometric parameters governing crease formation and self-folding, while the role of phase-transformation-driven behavior during deployment and recovery will be addressed in future investigations.

In all simulations, a time-dependent thermal load was applied, cooling the structure from the stress-free fabrication temperature (873 K) to room temperature (293 K) and back at a rate of 58 K/s.

## 2.3. Self-folding mechanism

The self-folding behavior of the proposed waterbomb origami structure is achieved by replacing the conventional mountain and valley creases with bi-material SMA-based actuators. Previous studies on bistable SMA actuators have demonstrated that thermal mismatch between an austenitic NiTi layer and a metallic layer, such as molybdenum (Mo), generates residual stresses during cooling from the stress-free crystallization temperature (~600 °C) to room temperature [22,28]. These residual stresses induce out-of-plane bending of the actuator, which can be exploited to initiate folding.

When the NiTi undergoes a phase transformation from austenite to martensite, the accumulated stresses are partially accommodated by the formation of self-accommodated martensitic twin variants, resulting in a reduction of the actuator curvature. Upon subsequent heating and transformation back to the austenitic phase, the deformation is recovered. In the present study, however, only the residual stress generation in the first step i.e., during cooling from the crystallization temperature was considered for this study, as this mechanism is essential for establishing the initial crease pattern required for self-folding [22,28].

To generate the prescribed mountain and valley folds within the origami unit cell, the crease lines are replaced by SMA-based bi-material hinges. Mountain folds (Figure 2b, red regions) are realized using NiTi/Mo bilayers, whereas valley folds (Figure 2b. blue regions) are implemented using NiTi/SS316L bilayers. The selection of these material combinations is based on the relative magnitudes of their coefficients of thermal expansion (Table 1), which satisfy the relationship $\alpha_{Mo} < \alpha_{NiTi\ (Austenite)} < \alpha_{SS316L}$. Consequently, cooling from the stress-free fabrication temperature generates bending in opposite directions for the two bilayer configurations, enabling the formation of both mountain and valley folds.

It is important to note that the NiTi layer is positioned identically in both hinge configurations, while the metallic layer is varied according to the desired folding direction. This approach allows deterministic programming of the crease orientation through material selection alone, thereby providing a scalable strategy for the realization of self-folding origami structures at microscale dimensions.

## 2.4 Parameter optimization

Design variations on length, thickness, width and shape were conducted to separate single effects and varied towards maximum deflection of the actuators. The different variations also consider the interplay of the parameters in an optimization procedure. Unless stated otherwise, each strip was modelled as a cantilevered actuator, mechanically fixed at one end with all remaining boundaries free, so that the reported longitudinal and lateral displacements reflect bending about this fixed edge under the applied thermal load.

### 2.4.1 Length Optimization

A parametric study was conducted for $L_{strip}$ of 50, 75, and 100 µm to evaluate the influence of actuator length on the thermo-mechanical response. The simulations showed a consistent increase in deformation with increasing $L_{strip}$, with the 100 µm strip exhibiting the highest displacement across all thickness combinations (Figure 3). This behavior can be attributed to the increase in the effective bending arm length of the bimetallic actuator. Since the deformation arises from the mismatch in thermal expansion between the two layers, a longer strip allows the thermal mismatch to act over a greater longitudinal distance, resulting in larger transverse displacements under identical thermal loading conditions. [29]

Although increasing the strip length enhances the deformation response, it also affects the geometric characteristics of the unit cell. Ideally, an origami unit cell consists of a collection of panels divided by the hinges, and the width of the hinge (which is $L_{strip}$) is ideally equal to zero. To conserve these origami design principles, the $L_{strip}$ should be a good compromise that allows self-folding and also conforms to origami design principles at the same time.

Considering the target unit cell dimensions of approximately 1.5 mm × 1.5 mm, a $L_{strip}$ of 100 µm was identified as the most suitable configuration. This length provided the highest thermo-mechanical deformation while preserving sufficient area for the passive folding regions and maintaining the desired unit cell architecture. Consequently, a strip length of 100 µm was selected for all subsequent optimization studies.

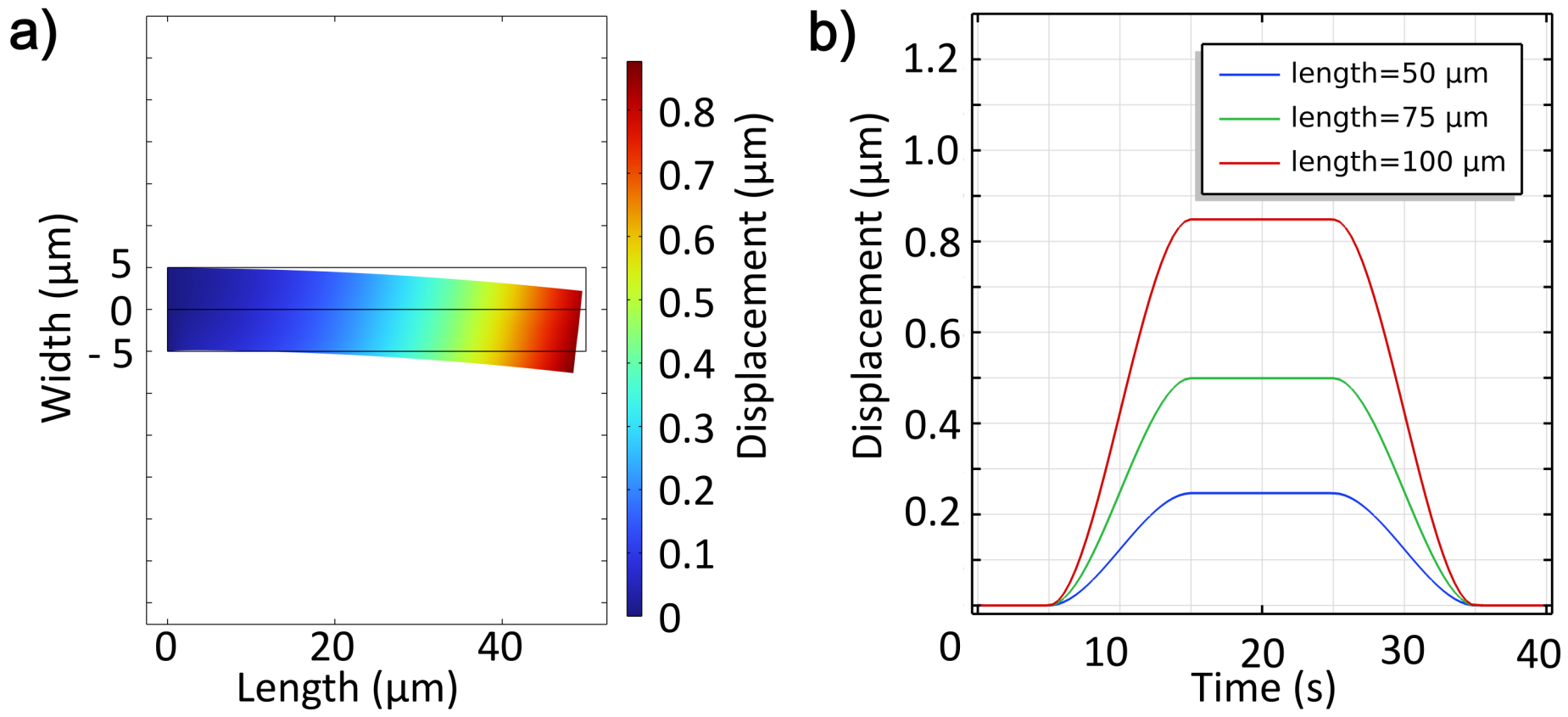

*Figure 3: Representative thermo-mechanical response during length optimization. (a) Displacement distribution in the bilayer strip during deformation. (b) Time-dependent displacement response of the actuator strips of varying lengths.*

### 2.4.2 Thickness Optimization

The thickness optimization study revealed that thinner layer combinations of ($T_{NiTi}$ / $T_{Mo}$) and ($T_{NiTi}$ / $T_{SS316L}$) generally produced larger thermo-mechanical deformation. Preliminary coarse sweeps showed that reducing both layer thicknesses (SMA and metallic layers) increased the displacement response (Figure 4a & c). However, the NiTi thickness was fixed at 5 µm, as thinner NiTi layers were found to compromise the structural integrity (experimentally) required for subsequent integration into the origami unit cell. Consequently, a secondary optimization focused on determining the optimum thickness of the top metallic layer (Mo and SS316L) for a fixed NiTi layer of 5 µm thick (Figure 4b & d).

Successive refinement of the parameter space resulted in optimum thickness combinations of 1.3 µm Mo–5 µm NiTi (Figure 4b, purple marker) for the mountain folds and 1.7 µm SS316L–5 µm NiTi (Figure 4d, purple marker) for the valley folds. This optimum was located through a multi-stage refinement, progressing from an initial coarse sweep (1–10 µm) down to a final increment of 0.1 µm for the metallic layer thickness once the region of interest was identified. These thickness combinations were subsequently used in all 3D and complete unit-cell simulations.

The observed behavior is consistent with classical Timoshenko bimetal strip theory [30] which states that the curvature developed in a bilayer structure is governed not only by the mismatch in the coefficients of thermal expansion, but also by the elastic moduli and thickness ratio of the two bonded layers. While reducing the thickness of the top metallic layer decreases the overall bending stiffness and promotes larger deformation, the resulting curvature does not increase indefinitely with decreasing thickness.

According to Timoshenko's theory, maximum curvature is obtained when a balance exists between the stiffness contributions of the two layers. Extremely thin top layers contribute very little to the bending moment generated by thermal expansion mismatch, whereas excessively thick layers increase the overall flexural rigidity of the bilayer and resist deformation. Therefore, an optimum thickness ratio exists rather than deformation increasing monotonically with decreasing thickness [30].

This behavior is consistent in both the NiTi/Mo and NiTi/SS316L systems, where the max displacement initially increases as the metallic layer thickness increases, reaches a maximum, subsequently levels off, and decreases. The optimum thicknesses of 1.3 µm for Mo and 1.7 µm for SS316L correspond to the point where the thermal mismatch-induced bending moment and the combined bending stiffness of the bilayer are most effectively balanced. These configurations should provide high actuation capability while maintaining sufficient mechanical robustness for integration into the self-folding origami-inspired unit-cell structure.

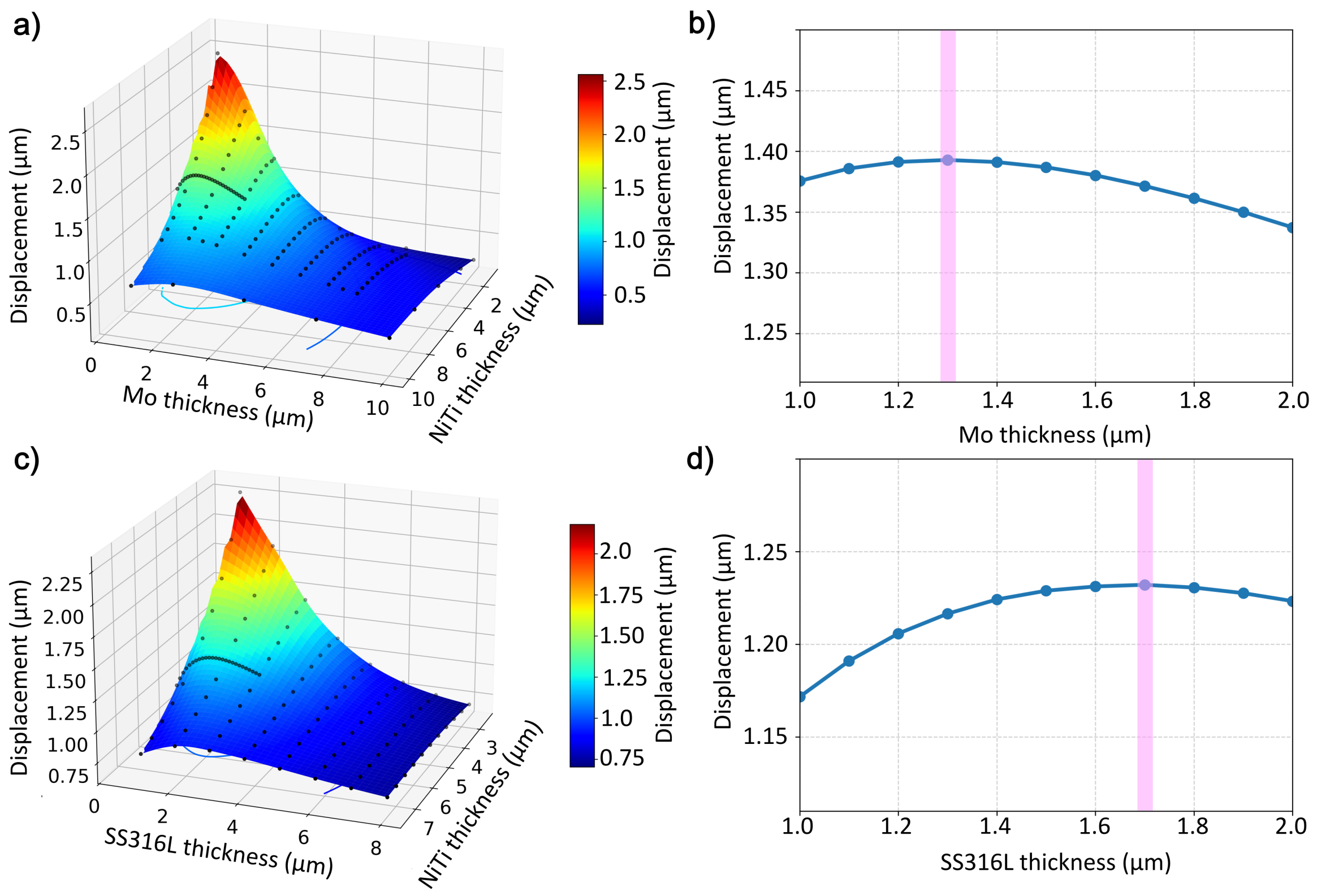

*Figure 4: Parametric sweep results for bilayer thickness optimization. (a) Influence of layer thickness combination of Mo and NiTi on displacement. (b) Influence of Mo thickness on displacement for a constant NiTi thickness of 5 µm. The total displacement increases with Mo thickness, has a max at 1.3 µm, and then decreases. (c) Influence of layer thickness combination of SS316L and NiTi on displacement magnitude. (d) Influence of SS316L thickness on displacement for a constant*

*NiTi thickness of 5 µm. The total displacement increases with SS316L thickness, has a max at 1.7 µm, and then slightly decreases.*

### 2.4.3 Width Optimization

Following the optimization of strip length ($L_{strip}$) and layer thicknesses ($T_{NiTi}$, $T_{Mo}$, $T_{SS316L}$), the model was extended to three dimensions to investigate the effect of strip width ($W_{strip}$) on the thermo-mechanical response (Figure 5a). Increasing the width produced a gradual increase in mean longitudinal displacement for both the Mo and SS316L actuators (Figure 5b, e), but also introduced undesired lateral bending and twisting across the actuator width. This lateral effect is captured by the difference (δ) between the corner and center displacement, which should be ideally zero (Figure 5c, f).

This trade-off — a modest gain in longitudinal displacement at the cost of growing lateral deformation with increasing width — motivated a ratio of the two quantities, with a higher ratio indicating better performance (Figure 5d, g). Widths below ~10 µm, though favorable by this metric, fall in a fabrication-constrained region (shaded purple) because the overhangs needed to deposit the additional metallic layers become impractical at that scale. Balancing actuation magnitude against deformation stability, a strip width of 15 µm was selected as the optimum: it delivers an average longitudinal deformation of ~1 µm while limiting lateral deformation to ~0.02 µm (Figure 5d, g, red markers).

This width-dependent lateral deformation can be understood through the limits of classical bimetallic strip theory. Timoshenko's theory assumes one-dimensional beam bending, under which curvature is independent of width and thermal stresses remain confined to a single bending plane, uniformly distributed across it [30]. Several studies have shown, however, that this assumption breaks down once the structure behaves as a plate rather than a beam [31–33].

The present three-dimensional geometry falls into this plate-like regime and is better described by Kirchhoff–Love plate theory, in which longitudinal bending couples with lateral deformation through Poisson effects. As strip width increases, thermal mismatch stresses are no longer confined to a single plane: additional transverse stresses develop and generate secondary bending moments across the width, producing lateral deformation and twisting [34–37]. This is consistent with the increasing δ observed with width for both actuators (Figure 5c, f): narrow strips deform almost uniformly, close to the ideal beam-like case (δ ≈ 0), while wider strips increasingly show anticlastic curvature and warping as the edges deform more than the center, so that the deformation is no longer purely unidirectional [38].

The observed width-dependence therefore does not contradict Timoshenko's classical theory, but reflects its known limits: a transition from one-dimensional beam behavior to two-dimensional plate behavior, in which transverse stresses and Poisson coupling become significant.

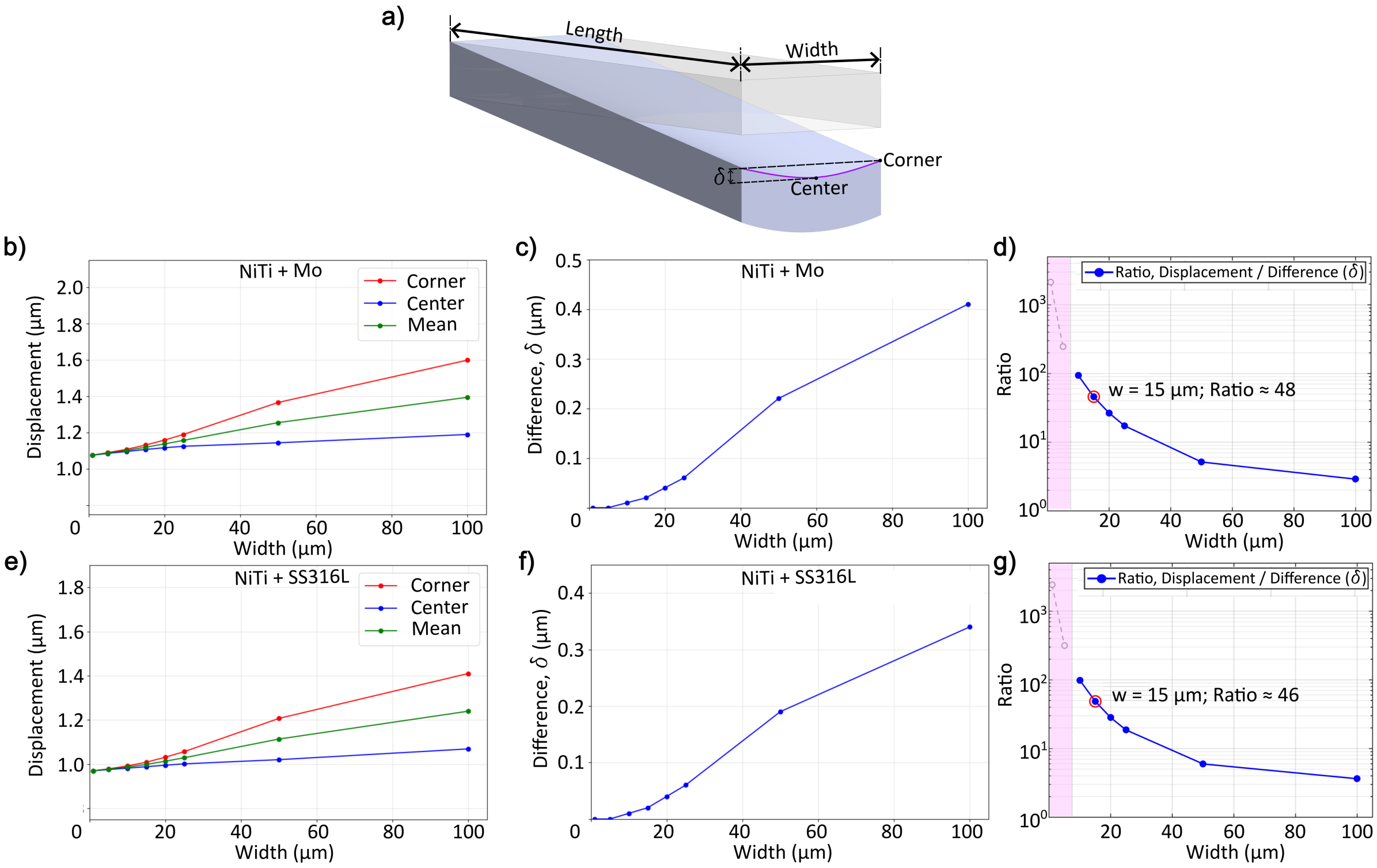

*Figure 5: Influence of strip width on deformation behavior. (a) Schematic of the 3D model showing longitudinal bending along the strip length (desired) and lateral bending across the width (undesired). A saddle-like deformation develops (purple line), with the corner-to-center deviation across the width defined as δ. (b, e) Longitudinal displacement as a function of actuator width for Mo and SS316L, respectively; the divergence between center and corner values reflects the unwanted lateral bending, which ideally would be zero. (c, f) Dependence of δ on actuator width for Mo and SS316L, respectively. (d, g) Ratio of longitudinal displacement (δ) as a function of width for Mo and SS316L, respectively; the purple-shaded region marks widths excluded by fabrication constraints (< 10 µm), and the red markers identifies the selected 15 µm width, where δ ≈ 0.02 µm.*

### 2.4.4 Shape Optimization

The shape optimization study was performed to investigate whether alternative strip configurations could provide improved actuation characteristics compared to the rectangular geometry. Figure 6 shows representative narrowing trapezoidal, rectangular, and widening trapezoidal geometries that were considered generated by parametrically varying the free-end width from 2 µm to 30 µm across twelve configurations, while the fixed end was held at a constant 15 µm.

While narrowing trapezoidal strips exhibited reduced deformation, an increase in tip width resulted in larger longitudinal displacement, with the maximum actuation observed for widening trapezoidal geometries. However, increasing the tip width also resulted in progressively larger unwanted lateral bending and twisting.

This behavior can be attributed to the redistribution of stiffness along the length of the strip. In narrowing geometries, the reduced cross-sectional area near the free end lowers the effective bending moment generated by the thermal mismatch and suppresses the overall deformation. In contrast, widening geometries increase the active area available for thermal expansion mismatch, resulting in larger actuation strokes. However, the asymmetric stiffness distribution causes the wider free end to become increasingly susceptible to lateral bending and out-of-plane warping [35].

From a plate mechanics perspective, wider tip regions exhibit lower resistance to torsional deformation and are more sensitive to anticlastic curvature arising from Poisson coupling. As a result, although widening geometries produce greater longitudinal displacement, they also introduce undesirable secondary deformation modes which compromise the directional precision of the actuator [35]. Therefore, the rectangular geometry corresponding to a constant width of 15 µm, as it provided the most favorable balance between actuation magnitude and unidirectional bending behavior.

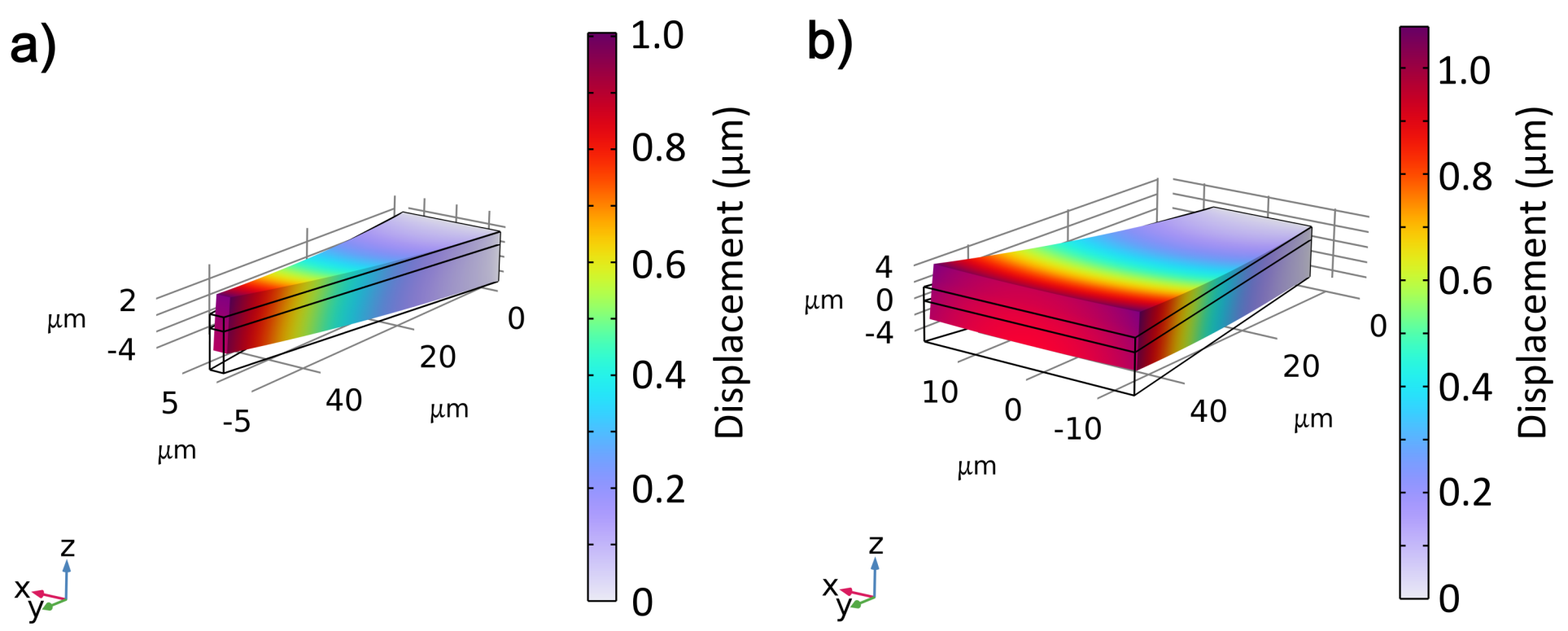

*Figure 6: Thermo-mechanical response of various strip geometries. In both cases, the fixed end was 15 µm. (a) Narrowing trapezoidal geometry with free end of 2 µm width. Shows a reduced displacement. (b) Widening trapezoidal geometry with free end of 30 µm width. Shows an increased displacement of the actuator.*

## 2.5 Unit Cell Integration and Full Geometry Model

The optimized strip geometries discussed in Section 3.3 were successfully integrated into the complete unit-cell model, resulting in coordinated thermo-mechanical deformation of the actuator strips and hinge regions (Figure 7 a–c). The resulting motion produced the intended self-folding mechanism of the waterbomb origami structure. The final displacement magnitude (~25 µm) is significantly higher than the maximum displacement obtained for the individual strip geometries (0.9–1.5 µm). This difference is mainly attributed to two factors: (1) the unit cell consists of an array of such strip geometries, which amplifies the overall deformation; and (2) the rigid motion suppression constraint assumed for the complete unit-cell simulation may also influence the total displacement achieved.

Despite this difference in magnitude, the deformation characteristics observed at the unit-cell level were found to be qualitatively consistent with those predicted from the individual strip studies, in terms of complete crease bending. This indicates that the overall response remains primarily governed by the local thermo-mechanical behavior of the actuator strips, and that the influence of interactions between neighboring strips and hinge regions is limited — that is, the deformation mechanism established at the strip level is preserved upon integration into the complete geometry. The observed bending from the initial flat state to a 3D configuration on cooling, following the provided crease pattern, should provide sufficient pre-folding to enable the complete fold to be achieved through subsequent crimping (manual folding).

Consequently, the bottom-up optimization approach adopted in this work provides a suitable framework for extending the design to larger tessellated arrays while maintaining similar local deformation characteristics.

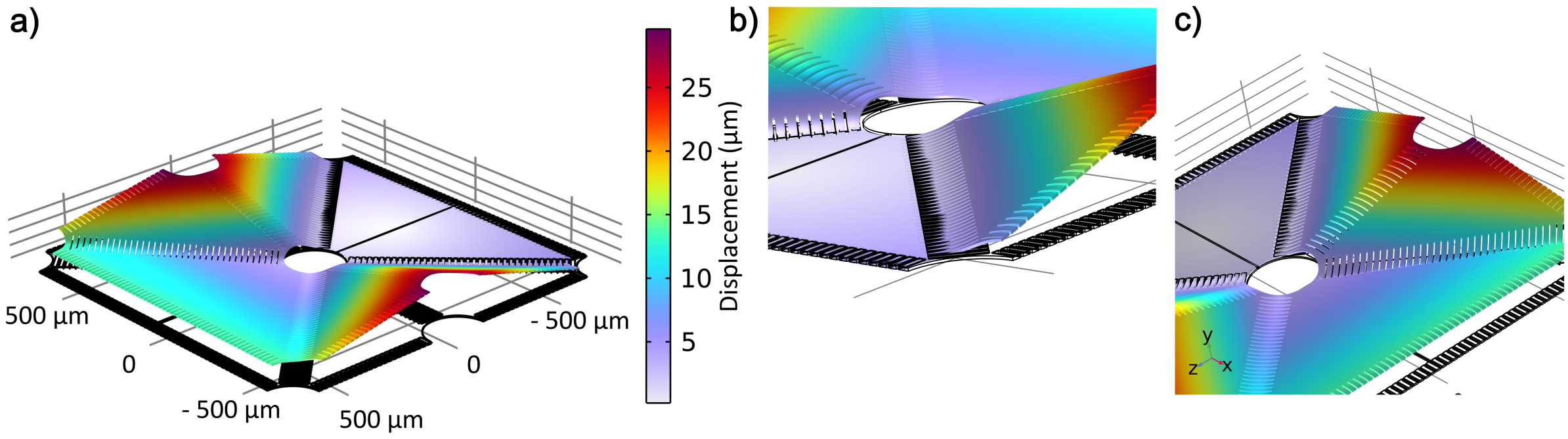

*Figure 7: Thermo-mechanical response of complete unit cell. (a) Both the undeformed and deformed complete unit cells. (b) Shows a close-up along a valley. (c) Shows a close-up along the hill hinge on the unit cell.*

## 2.6 Limitations

The current investigation has three main limitations. First, the assumption that NiTi behaves linearly is made to reduce the computational effort and is experimentally reasonable because the NiTi remains in the austenitic phase throughout the investigated temperature range. While this assumption allows us to capture the deformation caused by the thermal mismatch between the NiTi and the metal layers (Mo/SS316L), it does not accurately predict the resulting stress state within the materials.

Second, as a consequence of the first limitation, we cannot reliably determine the final stress state in either the NiTi or the integrated metal layers. Since stress concentrations may develop, greater care must therefore be taken during fabrication to achieve reliable and consistent results. The generated stresses may have different effects on the individual materials, and these effects should be considered when fabricating the actual origami devices.

Third, the assumptions made in the simulations do not account for variations in bonding between the different materials or for material-level irregularities. For example, the material properties can depend on factors such as layer thickness, grain size, and heat treatment.

Despite these limitations, the current results remain valid because they provide guidance within the design space for achieving miniature self-folding origami geometries. Since simulations inevitably deviate from experimental behavior, a more careful approach is adopted during the experimental fabrication and characterization to account for these limitations.

## 3. Conclusion

This work presents a simulation-driven design framework for the development of miniaturized self-folding origami structures based on SMA–based bi-material actuators. The proposed approach addresses a key challenge in the realization of origami-inspired neurovascular implants, namely the deterministic generation of mountain and valley crease patterns required for controlled self-folding at microscale dimensions.

Using coupled Multiphysics simulations, the influence of geometric design parameters on the self-folding behavior of a miniaturized waterbomb origami unit cell was systematically investigated. To enable efficient exploration of a large design space, NiTi was approximated as a linear elastic material, allowing the dominant thermal mismatch mechanism responsible for crease initiation to be captured while significantly reducing computational complexity. This approach facilitated the generation of an extensive parametric dataset that can serve as a design guideline for the optimization of self-folding origami structures.

Although the present study focuses on a waterbomb origami tessellation, the proposed methodology is not limited to this specific geometry. The developed framework can be readily extended to other origami tessellations and deployable architectures, enabling the realization of complex 3D structures tailored to patient-specific vascular geometries and aneurysm morphologies.

More broadly, the presented design strategy establishes a pathway toward next-generation self-folding implantable devices fabricated using microsystem technologies. By combining origami-inspired architectures with SMA-based actuation, it becomes possible to create deployable structures that contain localized low-stress regions suitable for the integration of sensing, communication, or feedback electronics. Such capabilities could enable the development of active neurovascular implants that not only provide structural support and flow modification but also offer real-time monitoring of device performance and treatment outcomes.

## Acknowledgements

The authors TS and PV acknowledge the funding received from DFG with the ‘Eigene Stelle’ research grant (VE1513/2-1).

## Data availability statement

The simulation data is available from the lead author on reasonable request.

## Conflict of Interest

The authors report no conflict of interest.

## Author contribution

SC: Methodology, Software, Validation, Formal analysis, Investigation, Writing – original draft, Visualization

TS: Resources, Supervision, funding acquisition, Writing – review & editing

PV: Conceptualization, Methodology, Supervision, Visualization, Writing – original draft, review& editing, Funding acquisition, Project administration.

# 4. Materials and methods

## 4.1 Material properties

Material properties for molybdenum were obtained directly from the COMSOL Multiphysics V6.2 material library. NiTi properties were included manually based on literature [39,40] and simplified as a linear elastic material as the material does not undergo phase change for the considered temperature change interval. SS316L properties were primarily sourced from the COMSOL Multiphysics V6.2 material library and supplemented with literature [41]. Listed in Table 1 are the material properties of Molybdenum, NiTi, and SS316L used for the simulations.

**Table 1:** Material properties used for thermo-mechanical simulations of the bimetallic actuator structures for self-folding origami.

| Property | Symbol | Molybdenum (Mo) | NiTi - Austenite (Linear Elastic Model) | SS316L |
|---|---|---|---|---|
| Density | ρ | 10,200 kg/m³ | 6,450 kg/m³[40] | 8,000 kg/m³ |
| Coefficient of Thermal Expansion | α | $4.8 \times 10^{-6}$ 1/K | $11.2 \times 10^{-6}$ 1/K[39] | $17 \times 10^{-6}$ 1/K |
| Young's Modulus | E | 312 GPa | 75 GPa[40] | 193 GPa |
| Poisson's Ratio | ν | 0.31 | 0.30 | 0.29 |
| Thermal Conductivity | k | 138 W/(m·K) | 11 W/(m·K)[39] | 18.9 W/(m·K) |
| Heat Capacity | Cp | 250 J/(kg·K) | 475 J/(kg·K)[39] | 500 J/(kg·K) |

## 4.2 Simulation framework

### Software and Modelling Framework.

The simulation workflow involved both COMSOL Multiphysics V6.2 (COMSOL AB, Stockholm, Sweden) and SolidWorks 2022 (Dassault Systèmes, Vélizy-Villacoublay, France). The geometries for the different optimization frameworks (strip length, layer thickness, strip width and shape) were modelled and analyzed directly in COMSOL, enabling rapid parametric sweeps and efficient optimization of the key geometric parameters. After the geometry and design parameters had been finalized, the complete origami unit-cell was modelled in SolidWorks and imported into COMSOL for the coupled thermo-mechanical analysis.

### Type of Study

A time dependent thermo-mechanical analysis was done to capture the transient deformation response of the actuator structures during the cooling and heating phase. The simulations were carried out by coupling the physics interfaces of solid mechanics and thermal expansion. This thermo-mechanical coupling enabled the model to undergo the temperature dependent deflection. The simulation framework consisted both 2D and 3D simulation frameworks, simplified 2D models were considered for the optimization of strip length ($L_{strip}$) and layer thicknesses ($T_{NiTi}$, $T_{Mo}$, and $T_{SS316L}$), allowing rapid parametric sweeps with reduced computational cost. Following the optimization of these parameters, the framework was extended to 3D models to investigate strip width ($W_{strip}$), shape optimization, inter-strip spacing, overhang dimensions, and the complete unit-cell behavior, thereby capturing width-dependent effects, lateral deformation, and the overall structural response.

## Governing Equations

The thermo-mechanical response of the actuator was modelled by coupling heat transfer, thermal expansion, and solid mechanics. The governing equations corresponding to each physics interface are presented below.

### *Heat Transfer*

The temperature distribution within the structure was obtained using the transient heat conduction equation together with Fourier's law:

$$\rho C_P \frac{dT}{dt} + \rho C_P \boldsymbol{u}.\nabla \mathrm{T} + \nabla.\mathbf{q} = \mathrm{Q} + Q_{ted} \quad (1)$$

$$q = -k\nabla T \quad (2)$$

where $\rho$ is the material density, $C_P$ is the specific heat capacity, $T$ denotes the temperature, $\boldsymbol{u}$ is the velocity field, $\mathbf{q}$ represents the heat flux vector, $k$ is the thermal conductivity, Q is the internal heat generation rate, and $Q_{ted}$ accounts for thermoelastic damping. Equations 1 & 2 govern the storage and conduction of thermal energy throughout the actuator and provide the temperature field required for thermo-mechanical coupling.

### *Thermal Expansion*

The temperature field was subsequently coupled to the structural analysis through the thermal strain relation

$$\varepsilon_{th} = \alpha(T)\left(T - T_{ref}\right) \quad (3)$$

where $\varepsilon_{th}$ denotes the thermal strain, $\alpha(T)$ is the coefficient of thermal expansion, $T$ is the current temperature, and $T_{ref}$ is the stress-free reference temperature. Equation 3 determines the strain induced by temperature variations and is responsible for generating mismatch strains within the bimetallic actuator.

### *Solid Mechanics*

The resulting stresses and deformations were obtained from the equations of linear elasticity.

$$0 = \nabla \cdot \boldsymbol{S} + \boldsymbol{f}_v \quad \mathbf{(4)}$$

$$\boldsymbol{\varepsilon} = \frac{1}{2}[(\nabla\boldsymbol{u})^T + \nabla\boldsymbol{u}] \quad (5)$$

$$S_{el} = \boldsymbol{C}:\varepsilon_{el} \quad (6)$$

where $\boldsymbol{S}$ is the Cauchy stress tensor, $\boldsymbol{f_v}$ represents the body force per unit volume, $\boldsymbol{\varepsilon}$ is the strain tensor, $\boldsymbol{u}$ is the displacement vector, $\boldsymbol{C}$ denotes the fourth-order elasticity tensor defined by Young's modulus and Poisson's ratio, and $\varepsilon_{el}$ is the elastic strain tensor. Equations 4 & 6 describe the equilibrium, deformation, and stress response of the structure.

The temperature field obtained from the heat transfer analysis is first converted into thermal strains through the thermal expansion relation. These strains subsequently act as inputs to the solid mechanics equations, where the resulting stress distribution and displacement field are determined. Consequently, the coupled formulation enables prediction of the thermo-mechanical deformation and deployment behavior of the hinges to obtain self-folding origami structures.

## 4.3 Boundary Conditions

### *Structural Boundary Conditions*

Structural boundary conditions were applied to constrain selected regions of the geometry while allowing the remaining region to deform. Based on the specific study the constraints of fixed and free boundary conditions considered on the actuators varied to mimic a meaningful scenario, and they are discussed in respective sections.

### *Thermal Boundary Conditions*

A time dependent temperature profile was applied uniformly to all boundaries of the structure, following the cooling-reheating cycle described in section 2.2 (873 K to 293 K and back, 58 K/s), consistently across all simulations.

### *Modelling Assumptions*

Throughout the simulation there were certain assumptions considered. Both the layers in the bimetallic actuator are perfectly bonded, all the materials exhibit linear elastic material behavior and all the material properties were considered to be temperature independent. NiTi was modelled as a linear material model as it did not undergo any phase transformations. Additionally, the bimetallic strip was assumed to be stress free at the initial configuration at 873 K.

## 4.4 Optimization framework

### Length Optimization

A parametric study was conducted by varying the strip length ($L_{strip}$) to investigate the effect of it on the thermo-mechanical response. To reduce the computational effort, symmetry was assumed and only half of the strip geometry was simulated. The simulated half lengths were selected as 25, 37.5 and 50 µm corresponding to full strip lengths of 50, 75 and 100 µm.

### Thickness Optimization

For mountain fold, NiTi-Mo-configuration, an initial coarse sweep was performed for thickness of 1, 2.5, 5, 7.5, and 10 µm for both NiTi and Mo. A refined sweep at 1–8 µm (1 µm increments) for Mo and 2.5–7.5 µm (0.5 µm increments) for NiTi; and a final sweep with NiTi fixed at 5 µm and Mo varied from 1–3 µm in 0.1 µm increments.

A very similar optimization strategy was applied for NiTi – SS316L configuration as well. For all the thickness optimization studies the $L_{strip}$ was fixed to a constant of 100 µm.

### Width Optimization

Following the optimization of the length and thickness the simulation framework was extended to 3D to investigate the influence of the strip width ($W_{strip}$) and capture the lateral deformation effects that cannot be captured in 2D.

The optimized values of $L_{strip}$ = 100 µm, $T_{NiTi}$ = 5 µm, $T_{Mo}$ = 1.3 µm, and $T_{SS316L}$ = 1.7 µm were retained and the same thermal loading conditions and equivalent boundary conditions of what were used in 2D were applied to the 3D models.

A parametric sweep study was conducted for strip width of 1, 5, 10, 15, 20, 25, 50, and 100 µm. The study evaluated both the desired longitudinal bending, which produces the out-of-plane folding required for hinge actuation, and the undesired lateral bending and twisting across the strip width. Lateral bending refers to transverse deformation perpendicular to the intended bending direction, resulting in warping or twisting of the actuator. Such deformation is undesirable because it reduces the unidirectional folding accuracy of the actuator and can compromise the coordinated deformation of adjacent strips within the origami unit cell.

### Shape Optimization

A shape optimization study was conducted to evaluate whether alternative shapes of the strip platforms could provide higher deformation compared to the rectangular geometry. The free-end width was varied as 2, 3, 5, 9, 12, 14, 15, 16, 18, 21, 25, and 30 µm, with fixed end at 15 µm, following the approach described in Section 2.4.4.

## 4.5 Unit Cell Integration and Full Geometry Model

Following the optimization of the individual strip parameters, the corresponding optimal dimensions and material combinations were incorporated into a representative complete origami unit-cell model to investigate the collective thermo-mechanical response of the unit cell.

A complete unit-cell geometry was modelled in SolidWorks (1.5 mm x 1.5 mm) using the optimized strip dimensions and subsequently imported into COMSOL for analysis. The SMA/Mo and SMA/SS316L actuators are placed along the hill and valley hinges of the unit cell, respectively. To prevent rigid body motion while allowing unrestricted deformation of the structure, rigid motion suppression was employed as the structural constraint. The same thermal loading and material assumptions used in the previous optimization studies were retained throughout the simulations.